\documentclass[preprint2]{aastex}

\shorttitle{Reconnection outflow}
\shortauthors{V\"{o}r\"{o}s et al.}

\begin{document}


\title{RECONNECTION OUTFLOW GENERATED TURBULENCE IN THE SOLAR WIND}


\author{Z. V\"{o}r\"{o}s$^1$,  Y. L. Sasunov$^1$,  V. S. Semenov$^2$, T. V. Zaqarashvili$^1$, R. Bruno$^3$, and M. Khodachenko$^{1,4}$ }
\affil{$^1$ Austrian Academy of Sciences, Space Research Institute,
    8042 Graz, Austria}
\email{zoltan.voeroes@oeaw.ac.at}
\affil{$^2$ Physical Institute, Saint Petersburg State University, St. Petersburg, Russia}
\affil{$^3$ INAF-IAPS, Istituto di Astrofisica e Planetologia Spaziali, Rome, Italy}
\affil{$^4$ Skobeltsyn Institute of Nuclear Physics, Moscow State University, Moscow, Russia}




\begin{abstract}
Petschek-type time-dependent reconnection (TDR) and quasi-stationary reconnection (QSR) models are considered to understand reconnection outflow structures and the features of the associated locally generated turbulence in the solar wind. We show that the outflow structures, such as discontinuites, Kelvin-Helmholtz (KH) unstable flux tubes or continuous space filling flows cannot be distinguished from one-point WIND measurements. In both models the reconnection outflows can generate more or less spatially extended turbulent boundary layers (TBDs). The structure of an unique extended reconnection outflow is investigated in detail. The analysis of spectral scalings and break locations show that reconnection outflows can control the local field and plasma conditions which may play in favor of one or another turbulent dissipation mechanisms  with their characteristic scales and wavenumbers.
\end{abstract}





\section{Introduction}

The reduced power spectral density (PSD) of turbulent magnetic fluctuations in the solar wind is known to follow power-law scaling $\sim f^{-\alpha}$ with spectral index close to $\alpha = 5/3$ (Kolmogorov scaling) or $\alpha = 3/2$ (Iroshnikov-Kraichnan scaling) over the inertial range of scales (see, e.g., Bruno \& Carbone 2005). Recently, much attention has been devoted to the turbulent heating of solar wind plasma, which occurs over characteristic ion/proton \citep{smith06, bou12} or/and electron scales \citep{alex09, chen13}, where the PSDs exhibit spectral breaks.

Large-scale velocity shears, for example regions of interacting fast and slow solar wind, can drive a turbulent cascade \citep{rob92}, influencing the proton heating rates through compressional effects \citep{staw11}.
In this Letter we are interested in reconnection outflows in the solar wind which generate velocity shears and interact with the background plasma in a complex way.
To study reconnection outflow associated structures we use WIND magnetic (fluxgate magnetometer, time resolution 3 s and 0.092 s, \cite{lep95}) and plasma data (time resolution 3 s and 92 s, \cite{lin95, ogi95}).
We study the database of reconnection outflow events compiled by \citet{phan09}.
A unique long-duration event from this database is considered in detail in Sections 2 and 3.
To explain the data we offer two Petschek-like reconnection scenarios with different reconnection outflow structures.
In Section 4, using the Morlet wavelets, we adaptively estimate the reduced PSDs for our selected events.
Discussions and conclusions are presented in the last fifth Section.

\section{Reconnection outflow structure}

We consider two Petschek-type reconnection scenarios: time-dependent reconnection (TDR) \citet{heyn96, sem04, sas12} and quasi-stationary reconnection (QSR) \citep{gosl05, phan09, gosl12}.

According to the TDR model (Figure 1, left) an unstable current sheet (tangential discontinuity - TD), via reconnection, decays into a system of moving large amplitude
fast, slow, Alfven and entropy waves. The waves propagate symmetrically outward from the reconnection site along the current sheet together with the reconnected plasma and magnetic flux, collectively
forming the reconnection outflow region with embedded discontinuities and a propagating flux tube of finite width. Observations of the structures in Figure 1 (left) depend on the geometry of crossings. Along the indicated trajectory (dashed magenta arrow), the border of a flux tube can be Kelvin-Helmholtz (KH) unstable TD \citep{sas12} generating turbulence behind the moving structure.

According to the QSR model (Figure 1, right) reconnection proceeds for a long time in a quasi-steady way producing a space filling wedge-shaped reconnection exhaust. Here the newly reconnected field lines maintained by the plasma inflow through the separatrix layer are strongly kinked. The kinks representing a pair of Alfv\'enic disturbances or rotational discontinuities (RD) in the inflow regions are accelerating plasmas to the exhaust from both sides as they propagate along the magnetic field away from the reconnection site. Since many exhausts contain plasmas with decreased magnetic field and enhanced proton density and temperature, the transition to he exhaust is often a slow-mode shock. It was found by \citet{hut07} that wave activity at the ion-acoustic range and near the electron plasma frequency is preferentially observed near the reconnection X lines or along the exhaust boundaries. These are the regions of large density and temperature gradients, anisotropies and shear flows.
\begin{figure}[t]
\includegraphics[width=80mm]{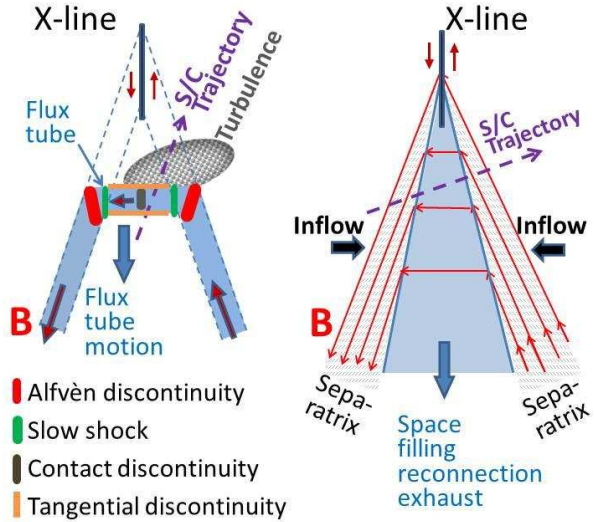}
\caption{\label{fig:event1}
Two-dimensional cartoons of Petschek-type reconnection outflow models. Left: Time-dependent reconnection (TDR). Right: Quasi-stationary reconnection (QSR).   }
\end{figure}

\begin{figure}[t]
\includegraphics[width=80mm]{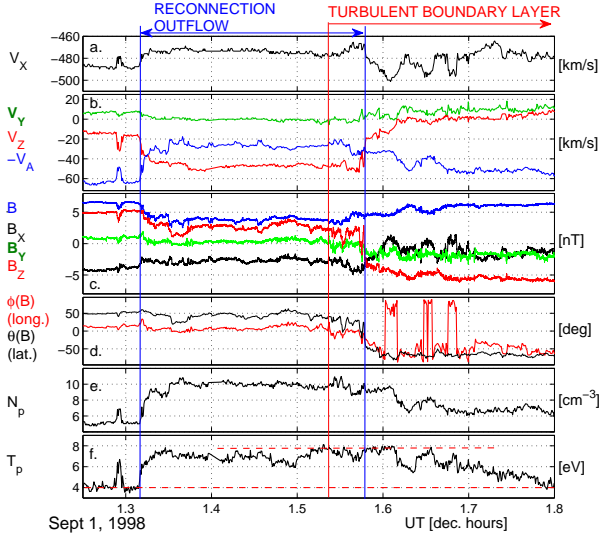}
\caption{\label{fig:event2}
Plasma and magnetic field observations (in the GSE coordinate system) during reconnection event on September 1, 1998.
(a.-b.) The bulk speed, its components and the negative Alfv\'en speed $-V_A$. (c.) Total magnetic field and its components. (d.) Longitudinal and latitudinal directional changes of the magnetic field vector; (e.) Proton density;
Proton temperature.}
\end{figure}

\begin{figure}[t]
\includegraphics[width=70mm]{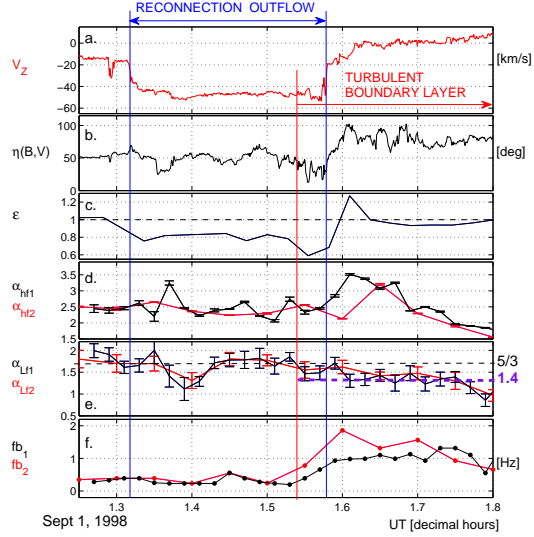}
\caption{\label{fig:event3}
Spectral characteristics of the event on September 1, 1998. (a.) $V_Z$ component of the bulk speed; (b.) Angle between $\mathbf{B}$ and $\mathbf{V}$; (c.) Anisotropy parameter $\epsilon = 1-\mu_0(P_{||}-P_\perp)/B^2 $; (d.-e.) High  ($\alpha_{hfi}$) and low frequency  ($\alpha_{Lfi}$) spectral indices computed in overlapping and non-overlapping sliding windows ($i=1,2$) of different lengths; (f.) Spectral break frequency.}
\end{figure}

Figure 2 shows a reconnection outflow event which occurred on September 1, 1998. Figures 2a,b show the bulk speed components and the local Alfv\'en speed.
Figures 2 c,d show the magnetic field magnitude, components, longitudinal and latitudinal directional changes of \textbf{B}. Figures 2e,f show the proton density and temperature, respectively. The reconnection outflow seen mainly in $V_Z$ is indicated by vertical blue lines. By using MHD stability criteria and analytical calculations it has already been suggested that for this event the outflow (a flux tube in TDR model) is bordered by TDs. The first TD at 1.31 UT was found to be KH stable.
However, when the RDs and slow shocks are merged, as may be the case at 1.31 UT, the signatures of TD are observed \citep{sas12} and the TDR (flux tube crossing) and QSR (exhaust boundary crossings) models cannot be distinguished.
In fact, at 1.31 UT (left border of the outflow) B decreases, both $N_p$ and $T_p$ increase, $\phi(B)$ and $\theta(B)$ show no significant directional changes.
The second TD at 1.59 UT (right border of the outflow) was suggested to be KH unstable by \citet{sas12}. Yet, due to the strong local fluctuations, it is impossible to distinguish between RD and TD without invoking the analytical model in \citet{sas12}. For this reason we do not assign any type of discontinuities to the outflow borders. Instead, we treat the whole interval from 1.54 UT (starting within the outflow, red vertical line) until $\sim$1.8 UT as a turbulent boundary layer. We suppose that the fluctuations within this boundary layer are driven by the predominantly vertical ($V_Z \sim$ -50 km/s) outflow rather than by normally occurring outflow-independent processes in the solar wind. This supposition is based on the following arguments: (a.) The fluctuations (e.g. seen in $V_X$, $B_X$) exhibit the largest amplitudes closer to the outflow border becoming weaker further away from their source; (b.) The highest proton temperatures are observed at the outflow boundary, then $T_p$ is slowly decreasing until 1.8 UT reaching the background level of solar wind proton temperatures before 1.3 UT (red vertical dashed lines in Figure 2f). Temperature fluctuations are correlated with density fluctuations. This indicates that the plasma is mixing and cooling within the boundary layer; (c.) There exist (anti)correlated fluctuations between speed, magnetic field, density and temperature fluctuations associated with strong azimuthal directional changes of the magnetic field $\phi(B)$ (Figure 2d). In the database of reconnection outflows \citep{phan09} similar changes in $\phi(B)$ occur frequently, often associated with both crossing of the outflow borders. In this event $\phi(B)$ reflects the change of the sign of $B_X$ (Figure 2c); (d.) The scaling properties of magnetic fluctuations are the same within the whole turbulent boundary layer (see later, Figures 3e,f).

Again, we are not able to differentiate between boundary layer fluctuations which can occur in TDR and QSR models. If the flux tube develops as it is suggested in the TDR model, the shear in $V_Z$ across the outflow boundary or TD is larger than the local Alfv\`en speed (Figure 2b) favoring the development of KH unstable boundary and turbulence. The strong directional changes $\phi(B)$ correlated with field and plasma fluctuations can indicate KH instability associated \citep{sas12} vorticity and vortex shedding ($V_Z \sim$ 0 km/s after 1.62 UT) in the wake of a moving flux tube ($V_Z \sim$ -50 km/s)\citep{grus10}. Although for strongly fluctuating $\phi(B)$ the occurrence of linear waves can be excluded, turbulence in the separatrix layer of the QSR model still can account for the observed changes as this is the region where the strongest fluctuation activity and instabilities can be expected \citep{hut07}. Changes in $\phi(B)$ can also correspond to crossing of turbulence generated structures \citep{grec09} or reconnection associated current sheet flappings \citep{voro09}. In any case, we suggest that the fluctuations are associated with the reconnection outflow. The next session is devoted to the analysis of the scaling features of fluctuations across the event in Figure 2.

\section{Nonstationarities and structures within reconnection outflow layers}
Contrarily to the commonly used length of solar wind data \citep{smith06, bou12, alex09, chen13} our intervals are rather short (see Figure 1) and extreme care is needed to identify the physical processes which can influence the scaling features.
To avoid spurious scaling estimations along the event crossing, we estimate parameters which can distort the reduced PSD. We consider the angle between the magnetic field and velocity vectors, $\eta(\mathbf{B},\mathbf{V})$, the pressure anisotropy parameter  $\epsilon = 1-\mu_0(P_{||}-P_\perp)/B^2 $ ($\mu_0$ - magnetic permeability, $P_{||}, P_\perp$ - pressures parallel and perpendicular to the magnetic field) \citep{liu12} and nonstationarity effects \citep{voro04, voro07}. Although the angle $\eta$ is computed from instantaneous values of  $\mathbf{B}$ and $\mathbf{V}$, it corresponds to $\eta(\langle\mathbf{B}\rangle ,\mathbf{V})$ ($\langle\mathbf{B}\rangle$ is the mean field) when locally the magnetic field direction is not changing. In such cases $\eta(\mathbf{B},\mathbf{V})$  controls the population of aligned or perpendicular fluctuations relative to the local magnetic field. The expected scaling indices are $\alpha_{||} = 2$ and $\alpha_\perp = 5/3$, respectively. It is found that roughly for $\eta(\mathbf{B},\mathbf{V})<30 ^o$ the contribution of field aligned fluctuations starts to be significant \citep{horb08}. Counterstreaming protons within reconnection outflows increase the proton pressure/temperature along the magnetic field \citep{liu12}, leading to $\epsilon \neq 1$, which can distort the spectrum near proton scales \citep{bale09}. Nonstationarity effects are recognized by estimating PSDs within two overlapping and non-overlapping sliding windows of different lengths (0.1 and 0.05 h).

Figure 3 shows the reconnection outflow in $V_Z$, the parameters $\epsilon$, $\eta$ and the estimated spectral features, explained below. The outflow borders and the turbulent boundary are indicated by vertical blue and red lines, respectively.
The PSDs are computed from the high resolution total magnetic field B. The scaling indices ($\alpha_{hfi}$) and ($\alpha_{Lfi}$), obtained from PSD fits over the high (subscripts ${hf}$) and low (subscripts ${Lf}$) frequency ranges, calculated within the two sliding windows (subscripts $i=1,2$), are shown in Figures 3d,e, depicted by black and red lines, respectively. The horizontal dashed lines in Figure 3e indicate the scaling indices 5/3 and 1.4, the latter corresponds to $\alpha_{Lfi}$ within the turbulent boundary layer. The frequency breaks ($fb_i$) estimated from PSDs, corresponding to the high frequency ends of low frequency fits within the windows $i$, are shown in Figure 3f.

Let us consider now the possible sources of nonstationarity. To minimize the effects of nonstationarity the fits were performed adaptively. The frequency ranges for the PSD fits were chosen so that the fits with the smallest errors were selected in each analyzing window. Figure 3d shows that the variations of $\alpha_{hfi}$ and the differences between the exponents for the two windows are much larger than the error bars (horizontal lines) of the fits. The largest differences between $\alpha_{hf1}$ and $\alpha_{hf2}$  occur when temporarily the angle $\eta(\mathbf{B},\mathbf{V})$ (Figure 3b) becomes less than 30 degrees, between 1.34 and 1.37 UT. Similar differences are seen between 1.57 and 1.62 UT when $\eta(\mathbf{B},\mathbf{V})$ changes from $\sim$25 to $\sim$ 90 degrees across the flow boundary. These changes in $\eta$ are associated with directional changes of the mean $\langle\mathbf{B}\rangle$ (not shown). Elsewhere the variations of $\langle\mathbf{B}\rangle$ are small.

Another source of nonstationarity is the temporal occurrence of pressure anisotropy. Figure 3c shows that the largest deviations in $\epsilon $ occur between 1.56 and 1.62 UT, overlapping in time with the largest changes in $\eta$. To further evaluate the impact of pressure anisotropies on magnetic fluctuations near the proton scale, we plot the proton temperature
anisotropy ratio $T_\perp/T_{||}$ versus $\beta_{||}$ (Figure 4). The curves indicate thresholds for mirror mode, proton cyclotron, parallel and oblique firehose instabilities \citep{hell06}. All the points which are inside the region bordered by instability thresholds correspond to stable situations with regard to temperature/pressure instabilities. The red and green points in Figure 4 are associated with the largest deviations of $\epsilon $ between 1.56 and 1.62 UT in Figure 3e.

Sudden jumps in the data, e.g. shocks, discontinuities, boundaries,  can also lead to spurious scalings with $\alpha \sim 2$. In fact, this is the case near the left border of the outflow, where both $\alpha_{Lf1}$ and $\alpha_{Lf2}$ show values near $2$, which are not associated with quasi-parallel population of $\eta<30 ^o$ (Figure 3b). However, between 1.25 and 1.42 UT we interpret the fluctuations in $\alpha_{Lfi}$ as a combined effect of the jump at the left border of the outflow (possibly TD) and $\eta<30 ^o$ within the analyzing windows. The frequency break points ($fb_i$) within the same time interval seem to be unaffected by nonstationarities (Figure 3f). Nevertheless, $fb_i$s changes from $\sim$ 0.3 Hz to $\sim$ 1 - 1.5 Hz when the spacecraft enters the turbulent boundary layer.

\section{Spatial scales of the spectral breaks}
The PSDs are estimated by excluding intervals containing sudden jumps, intervals of $\eta(\langle\mathbf{B}\rangle,\mathbf{V}) < 30 ^o$ or large deviations of $\epsilon$.
For the event on September 1, 1998 we found two time intervals, one within the reconnection outflow or flux tube, between 1.38 - 1.52 UT and one within the turbulent boundary layer region, between 1.62-1.76 UT. Within these intervals the reduced PSDs are
supposedly not distorted by other co-occurring physical processes. Repeating the same procedure we found one more long enough reconnection boundary layer crossing interval in the database of \citet{phan09}.
This event occurred on March 25, 1998, between 16.2 - 16.6 UT. The structure of the turbulent boundary layer is similar to the previous event (not shown).

Figure 5 shows the PSDs for both events. The bottom curve represents the only case for which the PSD is available within the space filling reconnection outflow (QSR model) or flux tube (TDR model).
The scaling exponents are determined through a least-square fit with an error  $\pm 0.1$ over the low frequencies and $\pm 0.03$ over the high frequencies. However, fluxgate magnetometers have limitations
over the high frequencies, where the signal to noise ratio becomes low.
Here we use a simple method to exclude the frequency ranges where noise dominates. The method is based on comparison of finding the frequency break using two different approaches. The intercept of low-frequency and high frequency power law fits should result in the same frequency break as the   high frequency end point of a well-defined low frequency fit. In Figure 5 the black circles correspond to the intercepts of two power-law fits, the red circles indicate the end points of the low frequency fits. The frequency ranges dominated by noise can be eliminated from a fit by trying to get the black and red circles closer to each other.

Figure 5 shows that the low frequency (inertial range) and the high frequency (kinetic scale) fits give scaling indices  $\alpha_{Lf}=1.4 - 1.7$  and $\alpha_{hf}=2.6 - 2.8$, respectively. These values are in agreement with previous studies in which much longer time series were analyzed in the solar wind \citep{alex09, bou12, smith06}. In the spacecraft frame, the frequency break point, within the turbulent boundary layer, ranges between $f_b$=0.8-1.5 Hz, which is different from $f_b$=0.4 Hz, observed within the reconnection outflow. In the pristine solar wind the break is observed at $f_b$=0.2-0.4 Hz \citep{bou12, smith12}.
\begin{figure}[t]
\includegraphics[width=70mm]{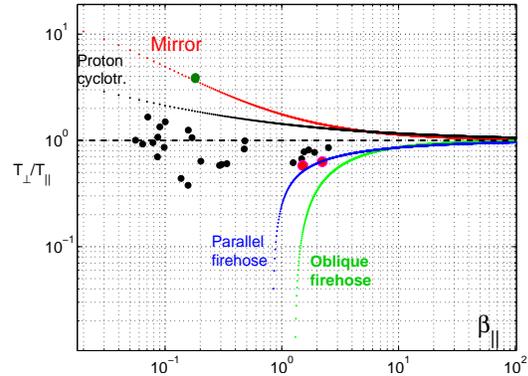}
\caption{\label{fig:event4}
Temperature anisotropy $T_\perp/T_{||}$ versus parallel plasma beta $\beta_{||}$ with instability thresholds for the event on September 1, 1998.}
\end{figure}

\begin{figure}[t]
\includegraphics[width=80mm]{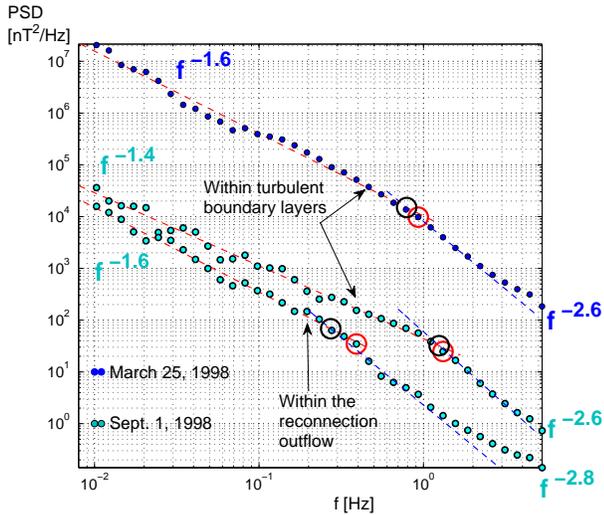}
\caption{\label{fig:event5}
PSDs  for two reconnection events. The black and red circles show the spectral break estimated with two different methods.}
\end{figure}

Now we calculate and compare the local wavenumbers at proton scales supposing frozen in fluctuations in the solar wind.
The wavenumber corresponding to the observed frequency break is $k_b=2\pi f_b/V$.
The wavenumbers corresponding to proton Larmor radius ($k_L=2\pi f_c/V_{th}$), inertial length ($k_i=2\pi f_p/c$) and resonant Alfv\'enic damping ($k_r=2\pi f_c/(V_A+V_{th})$) were also calculated. Here $V_A$, $V_{th}$ are the Alfven and thermal speeds, $c$ is the speed of light, $f_c$ and $f_p$ are the cyclotron and plasma frequencies, respectively.
On September 1, 1998, inside the outflow $k_b (outflow) = 0.0051$ $km^{-1}$, within the turbulent boundary layer  $k_b (TBL) = 0.015$ $km^{-1}$.
We found $k_b (outflow) = 0.0051 \sim k_r=0.0058$ $km^{-1} < k_L \sim k_i \sim 0.013$ and  $k_b (TBL) = 0.015 \sim k_L \sim k_i \sim 0.014 > k_r = 0.007$ $km^{-1}$.
For the event on March 25, 1998
$k_b (TBL) = 0.014 \sim k_L=0.018 \sim k_i=0.017 > k_r=0.01$ $km^{-1} $. The wavenumbers are determined by uncertainties of 10-15 \%.

\section{Discussion and conclusions}
We analyzed WIND data during crossings of reconnection outflow and outflow boundary structures in the solar wind. Two Petschek-type reconnection models were considered to explain the data: the time-dependent (TDR) and the quasi-stationary reconnection (QSR) models, leading to flux tube and space filling outflow structures, respectively (Figure 1). The models can explain the one-point measurements during crossing of an unperturbed flow boundary (merged RD and slow shock or TD) and outflow equally well. For the outflow boundary embedded in a strongly fluctuating field and plasma regions the TDR model predicts KH unstable TD with a developing turbulent vortex street behind the moving flux tube. For the same region the QSR model predicts intense wave activity, turbulence and instabilities within the outflow and separatrix layer. Although we could not distinguish between the two scenarios, we suggest that the fluctuations are driven by the reconnection outflow and form a unique turbulent boundary layer. The layer contains a slowly cooling-mixing plasma exhibiting a typical temperature-density profile with embedded significant directional changes of the magnetic field. The latter can correspond to turbulence generated structures, vortices or current sheet.

In the reconnection database of \citet{phan09} (51 events between November 1997 and January 2005) we found that 27 outflow events were accompanied by a boundary layer with embedded strong fluctuations.
However, only two crossing were long-enough for obtaining robust estimations of PSDs. Even if the crossing of turbulent regions were too short for the PSD calculations, these observations suggest that reconnection outflows can effectively generate fresh turbulence in the solar wind.

The PSDs indicate that within the outflow and turbulent boundary regions the observed scaling exponents over inertial and kinetic scales resemble those in the solar wind. Nonetheless, reconnection outflows can locally generate
turbulence with different locations of the spectral breaks associated with different characteristic wavenumbers. The changes of wavenumbers imply different dissipation mechanisms near proton/ion scales. For example, the proton inertial length is of the order of turbulence generated dissipation structures, current sheets \citep{dmi04}. The same spatial scale can be associated also by the Hall effect, which is able to steepen the PSDs \citep{gal06}. The Larmor radius can be associated with damping of kinetic Alfv\'en waves propagating at large angles relative to the local B \citep{leam98}. The wavenumbers associated with resonant Alfv\'enic damping were recently observed in several high velocity streams between 0.4 and 5 AU \citep{brun14}. Our results show that reconnection  can determine the particular local field and plasma conditions which may play in favor of one or another dissipation mechanism
in the turbulent solar wind. This supports the recent results of \citet{mark08} that the spectral break or the dissipation wavenumber cannot always be interpreted in terms of one single universal mechanism in the solar wind.

\acknowledgments
{\bf Acknowledgements} This work was supported by the Austrian "Fonds zur F\"{o}rderung der wissenschaftlichen Forschung" under projects P24740-N27 and S11606-N16. The work was also supported by EU collaborative project STORM - 313038. V.S.S. was supported by the Russian Science Foundation (RSCF) grant 243 14-17-00072, VVS.TZ was  supported by FP7-PEOPLE-2010-IRSES-269299 project- SOLSPANET, by Shota Rustaveli National Science Foundation grant DI/14/6-310/12 and by the Austrian "Fonds zur F\"{o}rderung der wissenschaftlichen Forschung" under project P26181-N27.  We acknowledge WIND spacecraft flux-gate magnetometer data from the Magnetic Field Investigation, plasma data from the
3D Plasma Analyser and from the Solar Wind Experiment. We acknowledge discussions with D. Burgess from Queen Mary University London.

\clearpage


\end{document}